# Phonon condensate in Landau-Pekar polarons and optical absorption due to their photoionization

E. N. Myasnikov[1], A.E. Myasnikova[2], and Z. P. Mastropas[1]
[1] Rostov State Pedagogical University, 344082 Rostov-on-Don, Russia
[2] Physics Department, Rostov State University, 344090 Rostov-on-Don, Russia

Quantum-field theory methods are applied to consider the state of the polarization field in a strongly-coupled large polaron (SCLP) and after its photodissociation. It is demonstrated that in the adiabatic approximation the polarization field in such a polaron coincides with the classical polarization field considered in Landau-Pekar theory up to small quantum fluctuations. However the state of this field after the polaron photodissociation is in principle different from that obtained in Pekar's and Emin's theories. This results from the fact that as it is shown below the polarization field in a SCLP is a spatially inhomogeneous phonon condensate. If the charge carrier is removed from the SCLP on its photodissociation the phonon condensate decays into phonons. As any Bose-condensate, the phonon condensate is a superposition of states with different number of quanta where summands are phased up to small quantum fluctuations. Therefore decay of the phonon condensate at the SCLP photodissociation results in an essentially wider band in the absorption spectrum than that predicted by theories with classical consideration of the polarization: the position of maximum is $4.2E_p$ and the half-width $\approx 2.5E_p$ ($3 E_p$ and $E_p$ in Emin theory, respectively), where $E_p$ is the polaron binding energy. The calculated band is in good conformity with mid-IR band in the optical conductivity spectra of complex oxides. The predicted ratio ($\approx 4$) of the maximum position of the band caused by the SCLP photoionization to the frequency of maximum of the band caused by phototransitions into polaron excited states calculated by other group is also in good conformity with experiments.

## 1. Introduction

Optical spectrum is an important criterion of applicability of a model to a description of a phenomenon. In the theory of a SCLP the optical absorption spectrum is discussed in the body of works. Pekar [1] has predicted that this spectrum consists of two parts: discrete lines caused by transitions of the charge carrier into excited states inside the polarization potential well and a continuum band due to the polaron photodissociation. He obtained an approximate position of the most intensive line ($1.3 E_p$) and a position of the low-frequency edge of the continuum band ($3 E_p$) where $E_p$ is the polaron binding energy. The value of energy necessary for the SCLP photodissociation ($3 E_p$) is negative of the electron energy in the ground state which is the sum of energy of the electron interaction with the polarization field and of the kinetic energy of the electron in a resting polaron. Pekar also pointed out that the energy necessary for photodissociation of a SCLP is three times larger than the energy necessary for its thermal dissociation because in case of thermal dissociation a crystal is depolarized by fluctuations and the electron is brought into the conduction band. During the SCLP photodissociation the polarization field and its energy ($2E_p$ in the classical case) does not change (according to Franck-Condon principle).

Then theoretical studies of two different parts of the polaron optical absorption spectrum were developed in different works. Transitions of the charge carrier into excited states in the polarization potential well were studied, e.g., in works [2-4]. Feynman, Hellwarth, Iddings and Platzman [2] have obtained impedance of a system of large polarons with the path-integral method. Devreese et al. [3] considered excitations of the charge carrier inside the polarization potential well taking into account that the excited state is relaxed (i.e. the phonon subsystem is adjusted to the excited state of the charge carrier). Mishchenko et al. [4,5] applying quantum Monte-Carlo method showed that sharp narrow bands predicted by Devreese et al. [3] would be broaden and form a single band. Its maximum is about $E_p$ in the intermediate and strong-coupling case [4,5].

Emin [6] has studied the optical absorption spectrum caused by the polaron photodissociation. He has predicted that the continuum band in the absorption spectrum caused by the SCLP photoionization starts at the energy $3E_p$, quickly achieves its maximum at energy about 3-3.5 $E_p$ and has a stretched high-energy tail. The width of the band was determined by the condition kR<1 where k is a wave vector and R is a polaron radius, i.e. the smaller the polaron is (in the frames of continuos approximation) the wider the absorption band is. Such an asymmetry of the band predicted by Emin has become a characteristic distinguishing feature between the small polaron case and large polaron case (for the small polaron the asymmetry of the optical absorption spectrum is opposite [6]). Emin [6] has considered the electron transition quantum-mechanically whereas the polarization field has been considered as a classical one like in Pekar work [1]. Therefore he has obtained the same value ($3E_p$) for the low-frequency edge of the continuum band as Pekar did.

However, the model with the classical consideration of the polarization field can provide information only on the average value of energy of the polarization field ($2E_p$). According to quantum theory each act of the polaron photoionization can be accompanied by radiation of random number of phonons but the probability of each event is determined in such a way that the average value of energy of the polarization field is $2E_p$. Therefore it is natural to expect a broader band due to the polaron photodissociation in the optical absorption spectrum than that predicted by theories with classical consideration of the polarization field. Similar broad bands were observed in the optical conductivity spectra of some complex oxides in the mid-infrared region [7-16]. Thus, the aim of the present article is to develop Emin consideration [6] by taking into account quantum character of the polarization field.

To do it we will use the results of the previous work [17] where we have shown that the polarization field in a SCLP is in a quantum-coherent state. In this state the fluctuations of the polarization field harmonics occur with respect to non-zero average values. These shifts of the equilibrium position in each harmonic of the polarization field caused by electron-phonon interaction are the parameters of the quantum-coherent state. They are determined with the minimization of the system free energy functional. Thus, the quantum-coherent states method in the frames of the quantum theory of the polarization field enables us to distinguish the coherent part of the polarization field. It is the part, which, in essence, is the classical polarization considered in works [1,6].

To proceed further let us first recall some ideas of the quantum-coherent state theory. As it is known, any Bose field in quantum physics can have average values of the field function different from zero [18]. Such a state of the field is called a coherent state [19,20], or a state with the deformation of the vacuum, or a state with the condensate of quanta [21,22] (or phonons in the case of the phonon field). Apart from this the field can exist in a state with the certain number of quanta. In such a state the average value of the field function is equal to zero. This property can be easily demonstrated in the example of a single quantum harmonic oscillator that can be one of the harmonics of any Bose-field. There exists an uncertainty relation which states that if the uncertainty of the number of quanta is equal to zero ($\Delta n = 0$) then the phase of the oscillator vibrations is completely uncertain [23], and, hence, quantum average values of the momentum and coordinate are equal to zero. In systems containing a phonon condensate $\Delta n \neq 0$. Non-zero quantum average values are characteristic for quantum-coherent states [19,20] with different degree of coherence. The more the degree of coherence is the more the uncertainty of the number of quanta is.

Quantum-coherent states can appear under the influence of classical devices or fields (e.g., radiation of a transmitting aerial) or due to the change of the degree of coherence of another quantum subsystem. For example, an electron in a stationary state of finite motion in a hydrogen-like atom obeys a certain degree of coherence. Indeed, an average value of potential of its electric field $<nlms_z | \alpha / | \mathbf{r} - \mathbf{R} | | nlms_z>$ (where $\mathbf{r}$ is an electron radius-vector) is in general different from zero in any point with the radius vector $\mathbf{R}$. On the contrary, the electron state with the certain momentum does not obey the coherence as

$$\lim_{L \to \infty} \int_{-L/2}^{L/2} dx\, dy\, dz \left| L^{-3/2} e^{i\mathbf{kr}} \right|^2 / |\mathbf{r} - \mathbf{R}| = 0.$$ Therefore, for instance, an electron transition from a

localized state in a medium into a state with the certain momentum changes the degree of coherence of its state. Simultaneously the degree of coherence of the polarization field generated by the electron changes too. A change of the degree of coherence of the electron field also occurs during its transition from one state into another in an impurity atom or molecule, or on a crystal defect, or in a molecule of a molecular crystal (molecular exciton). However the most interesting problem is the problem of the charge carrier transition from a SCLP state into a state with Bloch wave function as in such a case the changes in the coherent polarization field completely determine the electron binding energy.

It is shown in [17] that the SCLP includes the coherent lattice deformation. It is the coordinated in phases superposition of the lattice states with different number of phonons. For the phonon part of the SCLP its photodissociation is a quick Franck-Condon process [1,6], i.e. the phonon field does not change during the time of the phototransition. After the removal of the charge carrier from the SCLP during its photodissociation the coherent lattice deformation decays with the radiation of sufficiently large number of phonons in a single act. Indeed, according to the quantum-coherent states theory [19,20] a quantum-coherent state can be expanded in terms of states with the certain number of phonons. The expansion contains summands with all possible numbers of quanta from zero up to infinity, each summand with its weight. Therefore, a decay of the coherent part of the polarization field at the SCLP photodissociation is accompanied by the radiation of a random number of phonons in each event.

It should be stressed that this process can not be represented as successive acts of radiation of single phonons by the electron under the influence of the electron-phonon interaction. First, according to Franck-Condon principle the decay of the phonon condensate occurs without participation of the charge carrier, not during its transition into the free carrier state. Secondly, in each act of the single-phonon radiation by the electron due to electron-phonon interaction the initial momentum of the system is unambiguously distributed between the appearing electron and phonon in accordance with the momentum conservation law whereas in the case of simultaneous radiation of several phonons this distribution is indeterminate.

Let us note that separation of the polaron at the phototransition into two uncoupled parts can occur only in the case of strong electron-phonon coupling when $E_p >> \hbar\omega$, where $\omega$ is the phonon frequency. In the weak and intermediate-coupling case Franck-Condon principle is not applicable so that the electron transition into a new state after the photon absorption is accompanied by a change of the lattice state. Indeed, in this case the time of the phototransition is of the same order as the characteristic phonon time $\omega^{-1}$. Therefore, because of the photon absorption the weak/intermediate-coupling polaron passes again into a polaron state (a state of the electron coupled with phonons). I.e. the photodissociation of the polaron in the weak and intermediate coupling case cannot occur. As it is shown in [4,5] all the phototransitions between different polaron states lead to formation of a single absorption band. Its maximum is at about $E_p$ in the intermediate and strong-coupling case [4,5].

In the case of strong coupling polaron the photoionization can occur. At the photoionization the electron released by photon from the potential well becomes free and the state of the phonon field during this process does not change [6]. Therefore, in the strong-coupling case in agreement with the prediction of Pekar [1] there are two bands in the absorption spectrum. One band appears due to transitions into polaron states [4,5] and the other is due to the electron transition into the free carrier state in the conduction band (polaron photodissociation). As the polarization potential well in the polaron is close to Coulombic one [1], the intensity of both these bands is of the same order as it takes place for the phototransitions of the electron in the hydrogen atom into the states of discrete and continuous spectrum [24]. Such a peculiarity of the strong-coupling polaron is related with the spontaneous breaking of the translational symmetry of a crystal due to the SCLP formation [17].

Thus, although Frohlich polaron Hamiltonian describes only single-phonon processes at electronic transitions a large number of phonons appear simultaneously in a single act of the SCLP photoionization. The decay of the coherent part of the polarization into the random number of phonons in a single act of the polaron photodissociation allows to call it a phonon condensate with an uncertain number of quanta. Below we will show that the properties of the coherent part of the crystal polarization caused by the occurrence of a SCLP is analogous to Bose-condensate in a homogeneous system of non-interacting bosons at the temperature below the Bose-condensation temperature. But the reason of the appearance of the phonon condensate in a SCLP is not the peculiarity of boson quantum statistics but a coherence of the electric field of the electron when it is in a localized state. Therefore the phonon condensate is spatially inhomogeneous as it was pointed out in [17].

The article is organized as follows. First (Section 2) we recall the vector of state of the system of strongly coupled electron and phonon fields obtained in [17] with the quantum consideration of the latter. Then (in Section 3) an expression for the optical conductivity spectrum caused by the SCLP photodissociation is obtained. Section 4 demonstrates a good agreement between the theoretical spectrum calculated in the present work and mid-IR bands observed in experimental optical conductivity spectra of some complex oxides. Conclusion discusses the properties of the phonon field in the SCLP as those of the phonon condensate.

## 2. Quantum-coherent state of the polarization field in Landau-Pekar polaron

Let us first recall the vector of the ground state of a system of strongly coupled electron and phonon fields obtained in [17] with quantum-mechanical consideration of the polarization field. To study Landau-Pekar polaron [1] it is enough to consider a system which consists of one charge carrier and a field of longitudinal polar vibrations of the crystal lattice, corresponding to a single phonon branch, strongly interacting with the charge carrier. The interaction of the charge carrier with the polarization field is ordinarily described following Frohlich [25] in the second quantization representation. The polaron Hamiltonian has the form

$$\hat{H} = -\frac{\hbar^2}{2m^*}\nabla_r^2 + \sum_{k \neq 0}\left\{\hbar\omega_k b_k^+ b_k - \frac{e}{|\mathbf{k}|}\left(2\pi\hbar\omega_k V^{-1}\varepsilon^{*-1}\right)^{1/2}\left[b_k e^{i\mathbf{kr}} + b_k^+ e^{-i\mathbf{kr}}\right]\right\}, \quad (1)$$

where $b_k^+, b_k$ are the operators of creation and annihilation of a quantum of the k-th harmonic of the phonon field, $e$ is the charge of the carrier, $V$ is the crystal volume, $\varepsilon^* = \frac{\varepsilon_0 \varepsilon_\infty}{\varepsilon_0 - \varepsilon_\infty}$, $\varepsilon_0$ is a static dielectric constant, $\varepsilon_\infty$ is a dielectric permittivity at frequencies much higher than $\omega_k$.

In order to distinguish the coherent part of the phonon field a canonical transformation of Hamiltonian (1) and of the system wave function with the following unitary operator [19,20]

$$\hat{U} = \prod_k \hat{U}_k, \quad \hat{U}_k = \exp\{d_k b_k^+ - d_k^* b_k\}. \quad (2)$$

is used in [17]. Such an operator is ordinarily named a shift operator. It transforms the field operators $b_k^+, b_k$ into new ones $b'^+_k, b'_k$ according to a scheme

$$b'_k \equiv \hat{U}_k b_k \hat{U}_k^{-1} = b_k - d_k, \quad b'^+_k \equiv \hat{U}_k b_k^+ \hat{U}_k^{-1} = b_k^+ - d_k^*. \quad (3)$$

For the further consideration it is convenient to choose the new operators of the creation and annihilation of phonons in such a way that $\langle b'_k \rangle = \langle b'^+_k \rangle = 0$, and, consequently, $\langle b_k \rangle = d_k$, $\langle b_k^+ \rangle = d_k^*$.

In the case of strong electron-phonon coupling the adiabatic approximation is ordinarily used [1]. This allows us to seek the vector of the system ground state as a product of the electron and phonon parts:

$$|s\rangle = \psi_0(\mathbf{r} - \mathbf{R}, \beta)\exp\left\{\sum_{k \neq 0}\left(d_k b_k^+ - d_k^* b_k\right)\right\}|0\rangle, \quad (4)$$

where $|0\rangle$ is the vector of the ground state of the phonon subsystem in absence of the vacuum deformation, $\mathbf{R}$ is a real space vector, and $\psi_0(\mathbf{r}, \beta)$ is a normalized wave function of the electron in the polaron ground state with a localization parameter $\beta$ to be determined later. As it was shown in [17] in the zero-temperature limit the functional $\langle s|\hat{H}|s\rangle$ achieves its extremum with respect to the shift parameters $d_k = |d_k|e^{i\varphi_k}$ when

$$|d_k| = \frac{e}{|\mathbf{k}|}\sqrt{2\pi}\left(V\varepsilon^*\hbar\omega_k\right)^{-1/2}\eta_k(\beta), \quad (5)$$

$$\varphi_k = -\mathbf{kR} + 2\pi C(\mathbf{kR}), \quad (6)$$

where $C(\mathbf{kR})$ is an integer chosen in such a way that the phase $\varphi_\mathbf{k}$ belongs to the interval $[-\pi,+\pi]$. Let us note that, according to Exp.(6), the phase is expressed through the parameters which can have definite values. Hence, in a state $|s\rangle$ corresponding to the minimum of $\langle s|\hat{H}|s\rangle$ the polarization field

$$\langle \mathbf{P}(\mathbf{r})\rangle = i\sum_{\mathbf{k}\neq 0}\frac{\mathbf{k}}{|\mathbf{k}|}\hbar^{1/2}(2\mu V\omega_\mathbf{k})^{-1/2}\exp(i\mathbf{kr})(d_\mathbf{k}+d^*_{-\mathbf{k}}) \quad (7)$$

is quantum-coherent [19,20] and breaks the system translational symmetry [17].

In the extreme (5),(6)

$$\langle s|\hat{H}|s\rangle - \left\langle -\frac{h^2}{2m^*}\nabla^2_\mathbf{r}\right\rangle = -\sum_{\mathbf{k}\neq 0}2\pi e^2(V\varepsilon^*\mathbf{k}^2)^{-1}\eta^2_\mathbf{k}(\beta) = -\sum_{\mathbf{k}\neq 0}\hbar\omega_\mathbf{k}d^*_\mathbf{k}d_\mathbf{k}. \quad (8)$$

Exp.(8) demonstrates that we arrive at the same relation between the energy of the polarization field (here it can be called the energy of the phonon vacuum deformation) and the energy of interaction of the electron and phonon fields, as Pekar [1] did.

To minimize the functional $\langle s|\hat{H}|s\rangle$ in the parameter $\beta$ one should choose the form of a trial wave function $\psi_0(\mathbf{r},\beta)$ of the charge carrier. It is easy to see that if the coordinate dependence of this function is of a plane-wave type ($\psi_0(\mathbf{r})\sim\exp(i\mathbf{kr})$) then $\eta_\mathbf{k}$ is equal to zero. In such a case the deformation (5) and the energy (8) are equal to zero as well. Thus, choosing the vector (4) as the trial vector of the state, we limit ourselves to a consideration of such systems where the phonon vacuum deformation resulting in the break of the system translational symmetry is energetically profitable. As it is shown in Landau-Pekar works [1] such a break occurs in a system with strong electron-phonon coupling. In such a case Pekar trial wave function [1]

$$\psi_0(r,\beta) = (7\pi\beta^{-3})^{-1/2}(1+\beta r)\exp(-\beta r) \quad (9)$$

yields good results. Minimization of $\langle s|\hat{H}|s\rangle$ with respect to $\beta$ with the use of function (9) in neglecting the phonon frequency dispersion and the spatial dispersion of $\varepsilon^*$ yields the known Pekar values of the carrier binding energy in the ground state of the polaron:

$$E_p \equiv -\langle s|\hat{H}|s\rangle_{\min} = 0{,}054\frac{m^*}{m_e\varepsilon^{*2}}E_a = 0{,}108\alpha^2\hbar\omega, \quad (10)$$

where $\alpha = (e^4 m^*/2\varepsilon^{*2}\hbar^3\omega)^{1/2}$ is Frohlich constant of the electron-phonon coupling [25], and $E_a = m_e e^4/\hbar^2 = 27.2$ eV. The minimum corresponds to the value of $\beta$ obtained by Pekar [1]: 
$$\beta = m^*e^2/2\hbar^2\varepsilon^*. \quad (11)$$

The energy of the phonon vacuum deformation in this minimum obeys the relation
$$\hbar\omega\sum_\mathbf{k}d^*_\mathbf{k}d_\mathbf{k} = 2E_p = 0{,}216\alpha^2\hbar\omega. \quad (12)$$

Thus, the average energy of the phonon field in the quantum theory is the same as in the classical one.

### 3. Spectrum of the photodissociation of Landau-Pekar polarons at T=0K

The polaron photodissociation occurs as a result of interaction of an electromagnetic wave of frequency $\Omega$ with the charge carrier in the polaron (the longitudinal field of the polarization in the polaron obviously does not interact with the transverse electromagnetic wave). The operator of the interaction has the form

$$\hat{H}_{int} = \frac{e\hbar(\mathbf{kA})}{m^*c} e^{i\mathbf{Q}\cdot\mathbf{r}}, \qquad (13)$$

where $\hbar\mathbf{k}$ is the electron momentum, $\mathbf{A}$ is the amplitude of the vector potential of the electromagnetic field, related with its intensity $I$ as it follows: $I = \Omega\mathbf{A}^2/2\pi\hbar c$; $\mathbf{Q}$ is the wave vector of the electromagnetic wave. According to Fermi golden rule a probability of transition of the system from the state $|i\rangle$ into the state $|f\rangle$ per unit time under the influence of the operator $\hat{H}_{int}$ has the form

$$W_{if} = \frac{2\pi}{\hbar}\left|\langle f|\hat{H}_{int}|i\rangle\right|^2 \delta(E_i - E_f), \qquad (14)$$

where $E_i$ and $E_f$ are the energies of the initial and final states of the whole system. If the initial state is the ground state of the polaron in an electromagnetic field of a frequency $\Omega$ then

$$|i\rangle = \sqrt{\beta^3/7\pi}(1+\beta r)\exp(-\beta r)\prod_{\mathbf{q}}|d_{\mathbf{q}}\rangle \qquad (15)$$

and $E_i = -E_p + \hbar\Omega$.

To describe the final state we following Emin [6] use the simplest treatment of the photoelectric effect where the final state of the charge carrier is approximated as a free-carrier state. The vectors of possible final states of the phonon field are the eigen vectors $|\{v_{\mathbf{q}}\}\rangle = \prod_{\mathbf{q}}|v_{\mathbf{q}}\rangle$ of the non-shifted Hamiltonian $\hat{H}_{ph} = \sum_{\mathbf{q}}\hbar\omega_{\mathbf{q}}b_{\mathbf{q}}^+ b_{\mathbf{q}}$ describing the states with the certain number of quanta $v_{\mathbf{q}}$ in each harmonics. Thus, after the photodissociation the state (26) transforms into a state

$$|f\rangle = L^{-3/2}\exp(i\mathbf{kr})\prod_{\mathbf{q}}|\{v_{\mathbf{q}}\}\rangle, \qquad (16)$$

provided the sum of $v_{\mathbf{q}}$ (taking values 0 or 1) from the set $\{v_{\mathbf{q}}\}$ yields a certain number $v$. Hence, the energy of the final state is $E_f = \frac{\hbar^2\mathbf{k}^2}{2m^*} + \hbar\omega v$, if we neglect the dependence of $\omega$ on $\mathbf{q}$. Thus,

$$\delta(E_i - E_f) = \delta\left(-E_p + \hbar\Omega - \frac{\hbar^2\mathbf{k}^2}{2m^*} - v\hbar\omega\right). \qquad (17)$$

As the operator $\hat{H}_{int}$ acts only on the electron variables, the matrix element of the transition has the form

$$\langle f|\hat{H}_{int}|i\rangle = \int d\mathbf{r} L^{-3/2}\exp(-i\mathbf{kr})\hat{H}_{int}\sqrt{\beta^3/7\pi}(1+\beta r)e^{-\beta r}\prod_{\mathbf{q}}\langle v_{\mathbf{q}}|d_{\mathbf{q}}\rangle. \qquad (18)$$

Naturally, it contains the scalar product of the vector of a coherent state of the phonon field by a vector of its state with the certain number of phonons. After carrying out the integration in

(18) the probability of the electron transition into a state with the wave vector with modulus $k$ and direction in a spatial angle $\sin\theta\, d\theta\, d\varphi$ has the form

$$dW_{\{v_\mathbf{q}\},\mathbf{k}} = \frac{2\pi}{\hbar}\left\{\frac{e\hbar(\mathbf{kA})}{m^*c}32\sqrt{\frac{\pi}{7\beta^3}}L^{-3/2}\left(1+\beta^{-2}|\mathbf{Q}-\mathbf{k}|^2\right)^{-3}\right\}^2 \cdot \prod_\mathbf{q}|\langle v_\mathbf{q}|d_\mathbf{q}\rangle|^2\, d\rho(\mathbf{k}), \quad (19)$$

where

$$d\rho(\mathbf{k}) = \frac{m^*L^3 k(\varepsilon)}{(2\pi)^3 \hbar^2}\sin\theta\, d\theta\, d\varphi \quad (20)$$

is the spectral density of the final carrier states [6]. Let zero value of the angle $\theta$ corresponds to the direction of the vector $\mathbf{Q}$ and zero value of the angle $\varphi$ corresponds to the plane containing the vectors $\mathbf{k}$ and $\mathbf{Q}$. Then $\mathbf{kA} = Ak\sin\theta\cos\varphi$. According to (17) the electron momentum $\hbar\mathbf{k}$ and energy $\varepsilon$ in the final state are related as follows:

$$\hbar k(\varepsilon) = \sqrt{2m^*\varepsilon} = \sqrt{2m^*(\hbar\Omega - E_p - v\hbar\omega)}. \quad (21)$$

According to the energy and the momentum conservation laws (Exp.(17) and $\mathbf{Q} = \mathbf{k} + \mathbf{q}_0$, where $\mathbf{q}_0$ is the wave vector of the phonon field after the polaron photoionization, $\mathbf{q}_0 = \sum_\mathbf{q}\mathbf{q}v_\mathbf{q}$) an experiment can measure only the probability (19) summarized over all possible sets $\{v_\mathbf{q}\}$ having the same values of $v = \sum_\mathbf{q} v_\mathbf{q}$ and $\mathbf{q}_0$. A ratio of such a sum to the intensity $I$ of the exciting light has the form:

$$\frac{dW(\Omega,k,v,\theta,\varphi)}{I d\theta\, d\varphi} = \frac{256 e^2 k^3 \sin^3\theta\cos^2\varphi}{7\pi m^*\Omega c^2\hbar\beta^3}\left\{1+\beta^{-2}(\mathbf{Q}-\mathbf{k})^2\right\}^{-6} \cdot \sum_{\{v_\mathbf{q}\}}^v \prod_\mathbf{q}|\langle v_\mathbf{q}|d_\mathbf{q}\rangle|^2 \quad (22)$$

Here symbol $v$ over $\Sigma$ denotes that the sum is carried out over the sets $\{v_\mathbf{q}\}$ satisfying the condition $\sum_\mathbf{q} v_\mathbf{q} = v$. Besides, there is not a set with $v = 0$ among the sets $\{v_\mathbf{q}\}$. Indeed, in such a case $\mathbf{q}_0 = 0$, hence, $\mathbf{k} = \mathbf{Q}$, and $\mathbf{kA} = \mathbf{QA} = 0$, i.e. the probability of a transition with appearance of such a set $\{v_\mathbf{q}\}$ is zero. Obviously, the left part in Exp.(22) except the sum

$$P_v = \sum_{\{v_\mathbf{q}\}}^v \prod_\mathbf{q}|\langle v_\mathbf{q}|d_\mathbf{q}\rangle|^2 \quad (23)$$

is similar to Emin result [6] with two differences: (i) the form of dependence of the free carrier momentum k on the exciting light energy is different (Exp.(21)); (ii) we use Pekar form of the charge carrier wave function in the polaron. The former difference and the presence of the sum $P_v$ result from quantum consideration of the polarization field.

Let us consider the transformation of the phonon cloud of the polaron after its sudden "ionization" described by the sum $P_v$ in Exp.(22). The form of the factor in Exp.(22) caused by the phonon field transformation corresponds to the fact that photoionization of the SCLP represents a sudden perturbation for the polarization field in SCLP. Indeed, SCLP occurs in the case of strong electron-phonon coupling when the adiabatic condition $\hbar\omega \ll E_p$ [1] is satisfied. The condition of applicability of Franck-Condon principle to the phonon field transformation at the SCLP photoionization is obviously the same.

To calculate the sum in Exp.(23) let us take into account that according to the theory of quantum harmonic oscillator [19,20]

$$|\langle\{v_{\mathbf{k}}\}|d\rangle|^2 = \prod_{\mathbf{k}}|\langle v_{\mathbf{k}}|d_{\mathbf{k}}\rangle|^2 = \prod_{\mathbf{k}}(v_{\mathbf{k}}!)^{-1}|d_{\mathbf{k}}|^{2v_{\mathbf{k}}}\exp\{-|d_{\mathbf{k}}|^2\}. \quad (24)$$

The average energy of the k-th harmonic of the phonon field after the SCLP photodissociation

$$\overline{E}_{\mathbf{k}} = \sum_{v_{\mathbf{k}}=1}^{\infty}\hbar\omega_{\mathbf{k}}v_{\mathbf{k}}|\langle v_{\mathbf{k}}|d_{\mathbf{k}}\rangle|^2 = \sum_{v_{\mathbf{k}}=1}^{\infty}\hbar\omega_{\mathbf{k}}\frac{|d_{\mathbf{k}}|^{2v_{\mathbf{k}}}}{(v_{\mathbf{k}}-1)!}\exp\{-|d_{\mathbf{k}}|^2\}, \quad (25)$$

where $|\langle v_{\mathbf{k}}|d_{\mathbf{k}}\rangle|^2$ is the probability of appearance of $v_{\mathbf{k}}$ quanta in the k-th harmonic as a result of the perturbation. The average energy of the phonon field radiated due to the polaron photodissociation is, obviously, $\overline{E} = \sum_{\mathbf{k}}\overline{E}_{\mathbf{k}}$. After summarizing over $v_{\mathbf{k}}$ and taking into account Exp.(12) one obtains

$$\overline{E} = \sum_{\mathbf{k}}\hbar\omega_{\mathbf{k}}d_{\mathbf{k}}^{*}d_{\mathbf{k}} = 2E_p. \quad (26)$$

If we neglect the dependence of the phonon frequency on its wave vector in Exp.(26), one can see that the value

$$\sum_{\mathbf{k}}|d_{\mathbf{k}}|^2 = \overline{E}/\hbar\omega = \overline{v} \quad (27)$$

is the average number of phonons appearing in the polarization field as a result of a sudden perturbation. According to (26)

$$\overline{v} = 2E_p(\hbar\omega)^{-1}. \quad (28)$$

Taking into account the adiabatic condition $\hbar\omega \ll E_p$ the average number of phonons in the polarization field appearing due to the SCLP photodissociation $\overline{v} \geq 7.77$ that corresponds to Frohlich parameter $\alpha \geq 6$.

The sum in Exp.(23) calculated in Appendix 1 has the following form:

$$P_v = \frac{\overline{v}^{v-1}}{(v-1)!}e^{-\overline{v}} \quad (29)$$

As it can be seen from (29) the value $P_v$ rises with increasing $v$ until $v > \overline{v}$, then it decreases. Vertical lines in Fig.1 demonstrate the dependence of $P_v$ on $v$ for $v \leq 10$ in case of $\overline{v} = 7.77$ ($\alpha=6$).

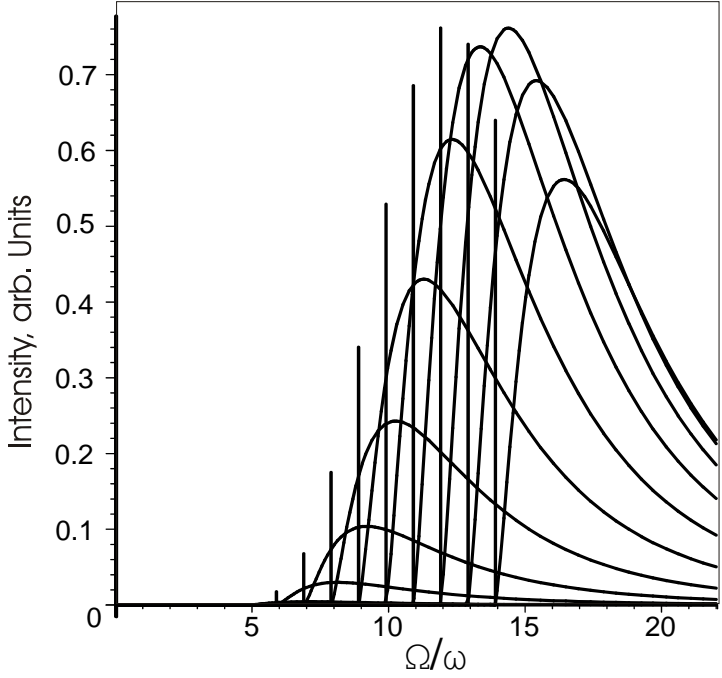

Fig.1. Vertical lines depict the probabilities of radiation of ν phonons (for $\nu \leq 10$) due to SCLP photodissociation at $\bar{\nu} = 2E_p(\hbar\omega)^{-1} = 7.77$ that corresponds to $\alpha = 6$. The solid curves are the partial spectra $P(\Omega,\nu,\theta,\varphi)$ of the SCLP photodissociation with radiation of ν phonons (for $1 \leq \nu \leq 10$) at $\bar{\nu} = 7.77$.

A factor $P(\varepsilon,\nu,\varphi)$ before $P_\nu$ in Exp.(22) determines the dependence of the probability of a carrier transition into a free state on its energy $\varepsilon = \hbar\Omega - E_p - \nu\hbar\omega > 0$ and on angles $\theta$ and $\varphi$. When a wavelength of the electromagnetic wave is much greater than the polaron radius, $\beta^{-2}\mathbf{Q}^2 \ll 1$. In fact, the probability (22), which can be represented as a product $P_\nu P(\Omega,\nu,\theta,\varphi)$, depends essentially on ν according to the factor $P_\nu$ and depends on the other variables according to the function $P(\Omega,\nu,\theta,\varphi)$.

The curves in Fig.1 demonstrate Exp.(22) as a function of Ω (with taking into account Exp.(21)) for different values of ν. Each curve begins at the energy $\hbar\Omega = E_p + \nu\hbar\omega$ and has a half-width of the order of $E_p$ and a stretched high-energy tail, whose length is determined by the bandwidth of the electron conduction band. The dependence of Exp.(22) on ν has the character of a shift of a curve to the value $\nu\hbar\omega$ (in the case of large values of ν the form of the curve also slightly changes). Therefore for given Ω and given direction of the free electron wave vector **k** one can observe electrons with various energies which differ to $\nu\hbar\omega$, where ν, according to Fig.1, changes from 1 up to approximately $2\bar{\nu}$. A curve with $\nu \approx \bar{\nu}$ begins at $\Omega = 3E_p/\hbar$ and has the same form as the light-absorption band due to polaron photodissociation obtained by Emin [6], but the integral intensity of the latter band is approximately $\bar{\nu}$ times higher.

The light-absorption spectrum (normalized to the number of polarons) is obtained by summing and integration of Exp.(22) (i.e. $P(\Omega,\theta,\varphi)P_\nu$) over ν and over $\theta$ and $\varphi$, respectively: $\sum_v \int dW(\Omega, v, \theta, \varphi)/I = \sum_v \int P(\Omega, \theta, \varphi) P_v d\theta d\varphi$. This spectrum is obviously a sum of partial absorption bands with given $v$ shown in Fig.1.

To compare the theoretical spectrum with experiments it is convenient to calculate the real part of the conductivity

$$\mathrm{Re}\,\sigma = \frac{\hbar \Omega N_p \sum_v W(\Omega, v)}{\varepsilon_\infty E^2} \qquad (30)$$

using Exp.(22) rewritten as a function of $\Omega$:

$$\mathrm{Re}\,\sigma = \frac{1024}{21} \frac{e^2 N_p}{m^* \Omega \sqrt{\varepsilon_\infty}} \sum_v \left[ 0.3\varepsilon(\Omega) \frac{\varepsilon^{*2}}{m^*/m_e} \right]^{3/2} \left[ 1 + 0.3\varepsilon(\Omega) \frac{\varepsilon^{*2}}{m^*/m_e} \right]^{-6} P_v,$$

or

$$\mathrm{Re}\,\sigma = \frac{1024}{21} \frac{e^2 N_p}{m^* \Omega \sqrt{\varepsilon_\infty}} \sum_v \left[ \varepsilon(\Omega) \frac{0.44}{E_p} \right]^{3/2} \left[ 1 + \varepsilon(\Omega) \frac{0.44}{E_p} \right]^{-6} P_v \qquad (31)$$

where $\varepsilon(\Omega)$ is determined by Exp.(21). Exps.(30,31) allows for the fact that polarons interact with the light in a medium with the refraction index $\sqrt{\varepsilon_\infty}$.

## 4. Results and comparison of them with those of Emin [6] and with experiments

The band in the optical conductivity spectrum due to SCLPs photodissociation calculated by Exp.(31) is shown in Fig.2 for α=6 (curve 1) and α=8 (curve 2) that corresponds to $E_p = 0.117$ eV and $E_p = 0.207$ eV, respectively, if the phonon energy is $\hbar\omega = 0.03$ eV. More precisely, Fig.2 demonstrates $\sigma(\Omega)/\left[\sqrt{\varepsilon_\infty}(m^*/m_e)\right]$ for a system with the polaron concentration $N_p = 10^{18}\,cm^{-3}$. Since Exp.(31) represents the optical conductivity in linear in the polaron number approximation, one can easily obtain the spectra for any other polaron concentration. As it is seen from Fig.2 the absorption band due to the SCLP photodissociation turns out to be unstructured. It is natural as we use the model where adiabatic condition ($E_p \gg \hbar\omega$) is satisfied and, hence, the distance ($\hbar\omega$) between the neighbor partial spectra (with the values of v that differ to unity) is essentially smaller than the half-bandwidth of a partial spectrum ($\approx E_p$ [6]).

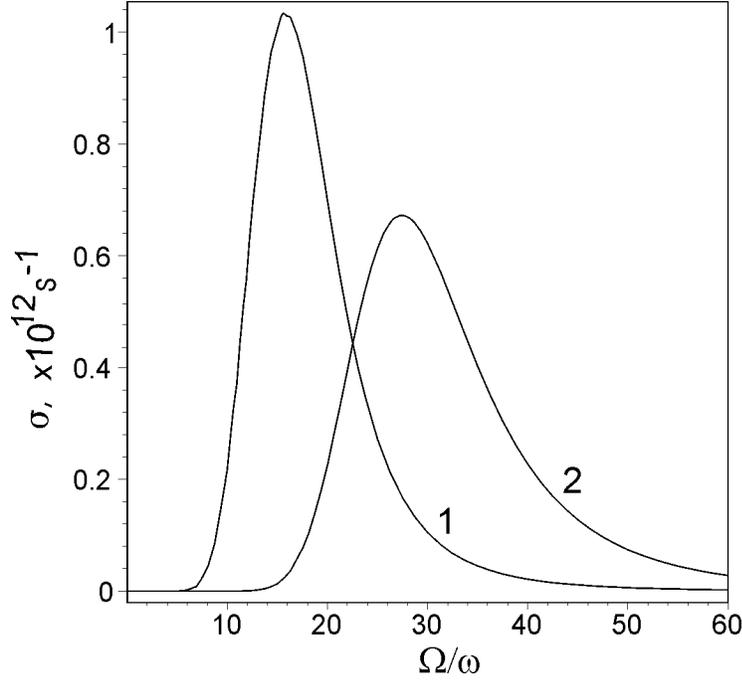

Fig.2. Optical conductivity $\sigma/(\sqrt{\varepsilon_\infty}\, m^*/m_e)$ caused by photodissociation of Landau-Pekar polarons at their concentration $N_p = 10^{18}\, cm^{-3}$. Curve 1 corresponds to α=6 ($\bar{v} = 7.776$, $E_p$ =0.1125 eV if the phonon energy $\hbar\omega$ =0.03 eV). Curve 2 corresponds to α=8 ($\bar{v} = 13.824$, $E_p$ =0.207 eV if $\hbar\omega$ =0.03 eV).

As it is seen from Fig.2 the optical conductivity band caused by the SCLPs photodissociation turns out to be wide with a maximum at $\Omega_{max} \approx 4.1 \div 4.2\, E_p/\hbar$, a half-width $\Delta\Omega \approx 2.2 \div 2.8 E_p/\hbar$ (in case of α=6-8) and a long-wavelength edge $\Omega_{edge} = E_p/\hbar + \omega$, as $v \geq 1$. Increase in the polaron binding energy results in an increase of $\Omega_{max}$ and $\Delta\Omega$ (curve 2). However, being expressed in units of $E_p/\hbar$ the value $\Omega_{max}$ remains unchanged: $\Omega_{max} \approx 4.2\, E_p/\hbar$ whereas $\Delta\Omega$ even decreases from 2.8 $E_p/\hbar$ to 2.2 $E_p/\hbar$ at the increase of $E_p$. Thus, in the case of α=6-8 one can calculate the polaron binding energy from its absorption spectrum as $E_p = \hbar\Omega_{max}/4.2$. If to determine the polaron binding energy from the half-width of the absorption band one obtains an interval: $\hbar\Delta\Omega/2.8 \leq E_p \leq \hbar\Delta\Omega/2.2$ if the electron-phonon coupling constant 6<α<8. Calculation of the polaron binding energy from the long-wavelength edge of the spectrum as $E_p = \hbar\Omega_{edge} - \hbar\omega$ will be of less accuracy, since the expression for the polaron absorption spectrum is obtained here in a model with one phonon branch interacting with the carrier and in case of T=0K. Presence of several such branches and anti-Stokes processes can essentially change the position of the long-wavelength edge of the band.

Thus, the optical conductivity band caused by the SCLP photodissociation is a wide unstructured band with a single maximum at the frequency $\Omega_{max} \cong 4.2 E_p/\hbar$ and a half-bandwidth of the order of $2.2 \div 2.8 E_p/\hbar$ (in the case of α=6-8). This band is calculated in T=0K approximation, i.e. for resting or moving slowly (with the velocities considerably lower than the minimum phase velocity of phonons interacting with the carrier) Landau-Pekar polarons.

Fig.3 demonstrates the difference between the band due to SCLP photoionization predicted by Emin in a model with classical polarization field [6] and that calculated above with quantum-mechanical consideration of the polarization field. Curve 1 is obtained by Emin expression, curve 2 is calculated according to Exp. (31) but with the carrier wave function in the polaron used by Emin, curve 3 is calculated by Exp.(31) with Pekar wave function of the carrier in the polaron. As it is seen from Fig.3 the spectrum obtained in the present work differs essentially from that predicted by Emin [6]. They differ in the low-frequency edge position: $\hbar\Omega_{edge} = 3E_p$ in [6] and $\hbar\Omega_{edge} = E_p + \hbar\omega$ in the present work; in the band maximum position: $\hbar\Omega_{max} = 3 \div 3.5 E_p$ in [6] and $\hbar\Omega_{max} = 4.2 E_p$ in the present work; and in the half-bandwidth of the band: $\hbar\Delta\Omega \approx E_p$ in [6] and $\hbar\Delta\Omega = 2.2 \div 2.8 E_p$ in the present work. The band obtained in the present work is more symmetrical with a stretched low-frequency tail apart from the high-frequency one. The reason of the difference is the quantum-mechanical consideration of the polaron polarization field in the present work. Such a consideration leads to a conclusion that the polaron photoionization with radiation of only one phonon is possible although the probability of such a process is small. Radiation of a large, exceeding the value $2E_p/\hbar\omega$, number of phonons is possible too. Only the average number of phonons is determined by the value $2E_p/\hbar\omega$, whereas the classical consideration of the polarization field in the polaron yields the result that each act of the polaron photodissociation results in a radiation of the energy $2E_p$ of the polarization field.

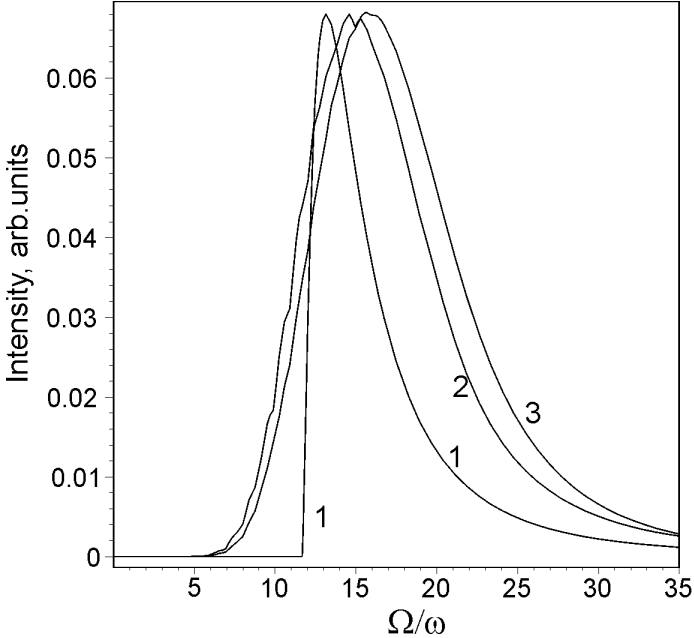

Fig.3. Light absorption caused by photoionization of SCLPs. Curve 1 is calculated according to [6], curve 2 and curve 3 are calculated by Exp.(31) with Emin and Pekar wave function of the charge carrier in the polaron, respectively. For all the curves $\bar{V} = 7.776$ (i.e. α=6), $E_p$=0.1125 eV, the phonon energy $\hbar\omega$=0.03 eV.

The discussed above form of the optical conductivity band represented in Fig.2,3 is like the form of the mid-infrared band observed in optical conductivity spectra of non-stoichiometric (doped to obtain free charge carriers) complex oxides: cuprates [7,8,14-16], nikelates [9,10], vanadates [11] titanates [12], manganites [13] and others. In all the experimental works it is stressed that the intensity of the mid-IR band is zero in case of zero doping and rises with an increase of doping. This demonstrates the relation of the origin of this band with the presence of free carriers. The other way to observe the mid-IR band is photoinduced light absorption in isolating parent compounds.

For example, works [9,10] represent the mid-IR band in the optical conductivity spectrum of $La_{2-x}Sr_xNiO_{4+\delta}$ (at T=300 K in [9] and at different temperatures from the very low ones up to the room temperature in [10]). The low-frequency edge of this band is at $E_{edge} \approx 0.1$ eV, and the interval of energies corresponding to increase of the absorption is about 0.4 eV. The maximum is at $E_{max} \approx 0.5$ eV, at higher energies the intensity slowly decreases. This band can be caused by a photodissociation of slow Landau-Pekar polarons with the binding energy about 0.12 eV. Practically the same mid-IR band as in $La_{2-x}Sr_xNiO_{4+\delta}$ with the maximum near 0.5 eV is observed in the optical conductivity spectrum of $La_{2-x}Sr_xCuO_{4-\delta}$ [7,8]. It also can be caused by photoionization of Landau-Pekar polaron with the binding energy about 0.12 eV.

Similar mid-IR bands were observed in other complex oxides. Work [14] reports the mid-IR optical conductivity spectra of $YBa_2Cu_3O_{6+y}$ (T=300K), $Nd_2CuO_{4-y}$ (T=10K), $La_2CuO_{4+y}$ (T=10K) and $La_{2-x}Sr_xCuO_{4+y}$ consisting of two peaks. Similar situation in $Nd_{2-x}Ce_xCuO_4$ is reported in [15]. Presence of two bands in the optical conductivity spectrum is in good conformity with the predictions of the SCLP theory. Indeed, in accordance with Pekar prediction [1] on the long-wavelength edge of the band caused by the SCLP photodissociation there should be a band caused by the charge carrier transitions from the ground polaron state into excited states in the polaron potential well. These bands were extensively studied in works [3,4,5]. In particular, for the strong-coupling case it is shown [5] that all these transitions result in a single band in the optical conductivity spectrum with the maximum at approximately $E_p$. In such a case the optical conductivity spectrum caused by Landau-Pekar polarons consists of two bands with approximate frequencies of the maximums $E_p$ and $4.2E_p$. The integral intensities of two these bands are of the same order since the form of the potential well in the SCLP is like that in the hydrogen atom where the oscillator strengths for transitions into excited states and for the photoionization are close [24].

The energies of maximum of the high-frequency bands caused by SCLPs photoionization are shown in the second column of Table 1. The third column contains the polaron binding energy estimated from the maximum position of the photoionization band as $E_p = \hbar\Omega_{photoioniz}/4.2$. The fourth column of Table 4 contains the energy $\hbar\Omega_{internal}$ of maximum of the band caused by the carrier transitions into internal excited states in the polaron potential well. In some works the low- and high-frequency bands are called d-band and mid-IR band, respectively [16]. As it is seen from Table 1 the energies of maximums of two bands in the optical conductivity spectrum of $YBa_2Cu_3O_{6+y}$ (T=300K), $Nd_2CuO_{4-y}$ (T=10K), $La_2CuO_{4+y}$ (T=10K) and $La_{2-x}Sr_xCuO_{4+y}$ are in the ratio $\Omega_{photoioniz}/\Omega_{internal} \cong 4 \div 4.5$ that is in good conformity with the theoretical prediction. Approximately the same ratio takes place between the maximums of two bands in $Nd_{2-x}Ce_xCuO_4$ [15].

Table 1

|  | $\hbar\Omega_{photoioniz}$, eV [14] | $E_p$, eV | $\hbar\Omega_{internal}$, eV [14] |
|---|---|---|---|
| $YBa_2Cu_3O_{6+y}$ | 0.62±0.05 | ≈0.155 | 0.16±0.03 |
| $Nd_2CuO_{4-y}$ | 0.76±0.01 | ≈0.17 | 0.162±0.005 |
| $La_2CuO_{4+y}$ | 0.6±0.02 | ≈0.14 | 0.13±0.02 |
| $La_{2-x}Sr_xCuO_{4+y}$ | 0.53±0.05 | ≈0.13 | 0.16±0.03 |

We calculated the photoionization band for the case of strong electron-phonon coupling $\alpha \geq 6$, that corresponds to $\overline{v} = 2E_p(\hbar\omega)^{-1} \geq 7.77$. The phonon energy corresponding to these parameters $\hbar\omega = 2E_p/\overline{v} \leq E_p/3.89$. It is in a good conformity with experimental data for these substances. For example, the phonon branches strongly interacting with the charge carrier in $Nd_{1.85}Ce_{0.15}CuO_4$ have the frequencies 115, 130, 146, 220 and 305 cm$^{-1}$ [26]. The energy of phonons strongly interacting with the charge carrier in copper-containing complex oxides were estimated on the base of experimental data in [27] as the value of the order of 0.02 eV.

The form of the mid-IR band in complex oxides is also in good conformity with that calculated by Exp.(31). It is demonstrated by Figs.4,5 where we fit the experimental optical conductivity spectra [14] with the use of Exp.(31). The width of the band calculated theoretically would be larger if we took into account the limitation of the wave vector space in the First Brillouin zone. Besides, we calculate the band for the case of medium with a single phonon branch strongly interacting with the charge carrier whereas in complex oxides there are several such branches ordinarily. Taking into account this fact will lead to a variety of the band forms. Allowing for the interaction of polarons will likely lead also to broadening of the band as it takes place in the weak-coupling case [28].

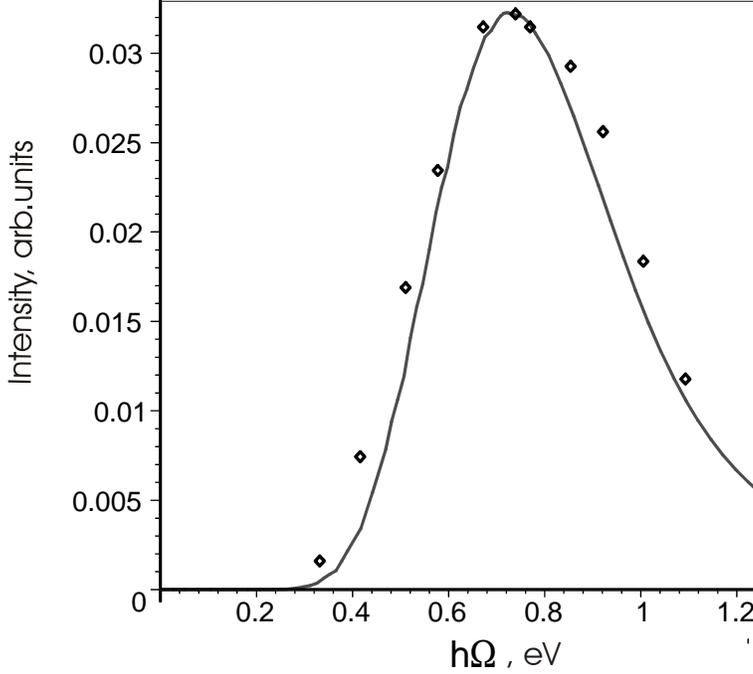

Fig.4. Mid-infrared band in the optical conductivity spectrum of $Nd_2CuO_{4-y}$ [14] at temperature T=10K (diamonds) and its fit by Exp.(31) with $\alpha$=6, $E_p$=0.18 eV (solid curve).

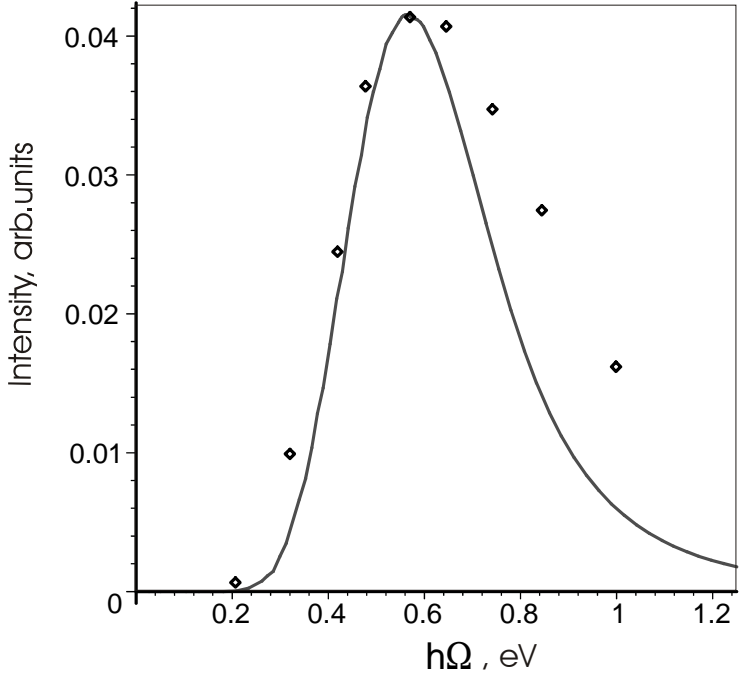

Fig.5. Mid-infrared band in the optical conductivity spectrum of $La_2CuO_{4+y}$ [14] at temperature T=10K (diamonds) and its fit by Exp.(31) with α=6, $E_p$ =0.14 eV (solid curve).

Authors of works [9] tried to fit the experimental mid-IR band in the framework of the small-polaron theory. They report absence of coincidence between the fit and the experiment for the frequencies above the mid-IR band maximum. The experiment unlike the fit contains an intensive high-frequency wing with the integral intensity comparable with the intensity in the main region of the band. In a spectrum calculated by Emin [6] and in the spectrum obtained in the present work such a wing is present. It appears as a consequence of a large bandwidth of the free electron band (often it is an order of magnitude larger than $E_p$ value).

The model of the small polaron theory is limited to consideration of narrow free carrier bands; their bandwidth should be much smaller than the binding energy of the polaron [29]. In works [9] the bandwidth of the free electron band is estimated (from the comparison of the band without the wing with the calculation according to the small polaron theory) as much smaller than the polaron binding energy. As a consequence the fit in the framework of the small polaron theory yields absence of a stretched high-frequency wing of the absorption spectrum that is ordinarily observed in experiments on complex oxides [7-16]. On the opposite, optical absorption spectrum due to SCLP photoionization contains such a stretched high-frequency wing. As it is seen from Figs. 4,5, the results of works [7-16] can be interpreted in the framework of Landau-Pekar polaron model.

We calculate the optical conductivity spectrum in the zero-temperature limit so that we cannot predict the temperature change of the band form and maximum position. However, we can predict the change of the integral intensity of the optical conductivity band caused by the SCLP photodissociation. The characteristic feature of the SCLP is the limitation of its velocity by the minimum phase velocity of phonons participating in its formation [30]. It results in a thermal destruction of the SCLPs when their thermal velocities exceed the minimum phase velocity of relevant phonons [31]. The thermal destruction expresses itself in a gradual decrease of the polaron concentration with temperature at critical temperatures much lower than $E_p$. (The critical temperature is determined by the values of the phonon minimum phase velocity and the polaron binding energy [31,32].) Therefore, the integral intensity of the band will decrease with the increase of temperature. The transition of the charge carriers from the polaron states into the free carrier states results also in lowering their effective mass. Hence, a corresponding decrease of the system resistance (the so-called activational behavior) should occur in the same temperature interval [32].

An example of the polaron thermal destruction at temperatures low in comparison with the polaron binding energy is represented by optical conductivity spectrum of $\beta - Na_{0.33}V_2O_5$ [11]. It contains a mid-IR band with the maximum at 3000 $cm^{-1}$ that can be caused by photodissociation of SCLP with the binding energy about 0.09 eV. As it is shown in Fig.6, the form of the band in the low-temperature spectrum (T=5K) is in good conformity with its fit by Exp.(31) with $E_p$=0.089 eV, α=6. The phonon energy corresponding to these parameters $\hbar\omega = 2E_p / \bar{v} = 0.025$ eV that is in good conformity with the reflectivity spectra [11], demonstrating large longitudinal-transverse splitting (i.e. strong electron-phonon interaction) for a set of phonon branches with the frequencies 80-300 $cm^{-1}$. Already the optical conductivity spectrum of $\beta - Na_{0.33}V_2O_5$ [11] corresponding to T=145K demonstrates decrease of the integral intensity of the mid-IR band, and the 300K spectrum shows approximately two times lower integral intensity in comparison with that for T=5K spectrum. Simultaneous study of the system resistance would reveal its "activational" behavior due to decrease of the carrier effective mass at their transition from the polaron state into free carrier state.

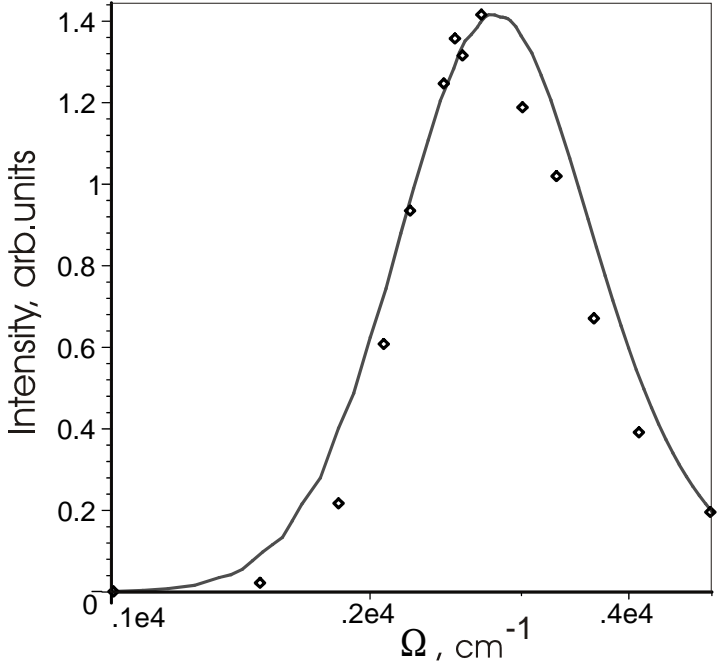

Fig.6. Optical conductivity spectrum of $\beta - Na_{0.33}V_2O_5$ [11] for E||b at temperature T=5K (diamonds) and its fit by Exp.(31) with α=6, $E_p$=0.089 eV (solid curve). The scale in the abscissa axis is logarithmic.

    For example, such an activational behavior of the resistance is observed in $LaTiO_{3.41}$ [12] in the same region of temperature where its optical conductivity spectrum demonstrates an essential decrease of intensity of the mid-IR band. Indeed, in the region of temperatures 150-300 K the integral intensity of the mid-IR band decreases 3-4 times. In the same temperature region the resistance along the direction ||a demonstrates change of its temperature behavior: in the region 50-150 K the resistance increases with the temperature increase whereas in the region 150-300 K the increase stops and the resistance remains practically constant with increasing temperature. It is in a good agreement with the temperature behavior of the resistance in systems with Landau-Pekar polarons [30-32]. The energy of the maximum of wide mid-IR band observed in E||a optical conductivity spectrum of $LaTiO_{3.41}$ is about 1700 $cm^{-1}$ at low temperature. Such a band can be caused by photodissociation of the SCLPs with the binding energy $E_p \approx 0.043$ eV. Of course, such an interpretation is possible only if among the phonon branches in $LaTiO_{3.41}$ there is a low-frequency branch strongly interacting with the charge carrier. For example, such a situation takes place in $SrTiO_3$ where the phonon branch 167 $cm^{-1}$ [33] interacts strongly with the charge carrier.

    It is useful to note that even in an atom a transition of electron from one level to another leads to a change of the coherent part of the carrier electric field. If such a transition occurs in an impurity atom or in an ion in a polarizable medium it must result in a radiation of quanta of the polarization vibrations. If Franck-Condon principle can be applied to this transition then a form of the corresponding absorption band should obey Exp.(31) without smearing out the lines forming the band due to possible different values of the electron energy in the final state. Work [34] represents an example of such light emission band caused by

electron transitions in impurity ion $Ni^{2+}$ in perovskite materials $KMgF_3$ и $BaLiF_3$. The form of the observed emission band (with its fine structure) is in a surprisingly good conformity with the structured band consisting of separate lines (showed by vertical lines in Fig.1) if to make a correction caused by smearing out of each line due to the phonon frequency dispersion. The difference of the electron binding energy in these two states caused by a difference in the polarization according to data of [34] can be estimated by the value of the order of 0.01 eV.

We would also like to note theoretical work [35], calculating the spectra of photodissociation of Jan-Teller polarons in $LaMnO_3$ and $CaMnO_3$ crystals. Without using the quantum-coherent theory methods the authors of [35] also predict appearance of wide bands albeit of symmetrical (Gaussian) form (i.e. without any stretched high-frequency wing) like the spectra of F-center ionization.

Influence of the temperature on the band caused by the polaron photodissociation will naturally lead to an increase of its width. However, the theory of photodissociation of Landau-Pekar polarons in case of non-zero temperatures is complicated essentially. Coherent polarization in Landau-Pekar polaron cannot propagate with a velocity higher than the phase velocity of phonons participating in the polaron formation [17,31]. In the case considered by Pekar and in the present work we use the model with dispersionless phonons that corresponds to zero phase velocity of phonons. In such a model one cannot consider thermal motion of polarons. Besides, the distribution function of polarons has peculiarities caused by a competition of polaron states of a charge carrier and their Bloch states. The problem of the optical conductivity spectra of polarons at finite temperatures will be considered elsewhere.

## 5. Conclusion

Thus, we have demonstrated that photodissociation of the SCLP being quick Franck-Condon process results in a separation of the polaron into two parts. The phonon structure of the polaron becomes free at the SCLP photodissociation and decays not under the influence of the electron-phonon interaction but independently. Since the number of phonons in the phonon structure of the SCLP is uncertain, it decays into different number of phonons in different events. This property is one of the characteristic features of a condensate.

Let us demonstrate that phonon "coat" of the SCLP being a quantum-coherent state of the crystal lattice has other characteristic features of the phonon condensate. Expansion of the coherent state in terms of states with the certain number of quanta [19]

$$|d_{\mathbf{k}}\rangle \equiv \exp\left(-|d_{\mathbf{k}}|^2/2\right) \sum_{n=0}^{\infty} \frac{|d_{\mathbf{k}}|^n \exp(in\varphi_{\mathbf{k}})}{(n!)^{1/2}} |n\rangle$$

shows that in any harmonic the summands with the number of phonons different by the unity differ in their phases by the same value $\varphi_{\mathbf{k}}$, as it takes place in the wave function of Bose-condensate [20]. (Let us recall that according to Exp.(6) $\varphi_{\mathbf{k}}$ is a definite, non-fluctuating value.) Therefore, we can call the system of phonons coupled with a charge carrier in the SCLP a condensate. Contrary to Bose-condensate of Cooper pairs in conventional superconductors, characterized by accumulation of bosons in some region of the momentum space, the condensation of phonons in a SCLP occurs in a vicinity of center of the charge carrier localization in the coordinate space. Indeed, the shifts $d_{\mathbf{k}}$ in different harmonics are phased in a corresponding way. Namely, according to Exp.(6) the difference between $\varphi_\mathbf{k}$ values in neighboring harmonics is determined by the difference between their wave vectors **k** which does not depend on **k**.

The phonon condensate in a SCLP decays at not only the SCLP photodissociation. It disappears also when the polaron velocity becomes exceeding the minimum phase velocity of relevant phonons [17]. The SCLP moving with a velocity exceeding the minimum phase velocity of one of phonon branches interacting with the charge carrier generates coherent phonon radiation similar to Cherenkov radiation of electromagnetic wave [17,31]. This coherent polarization wave is also a phonon condensate but of the other form, similar to the photon condensate generated by laser. As work [17] demonstrates it is the polaron braking by the coherent phonon radiation, which was calculated by Thornber and Feynman [36]. As it was shown in Landau Bose-liquid theory [37] existence of a similar critical velocity is characteristic for Bose-condensate in a superfluid helium. This is an additional evidence of the fact that the phonon "coat" of Landau-Pekar polaron is a condensate of phonons. This conclusion allows to state that Bose-condensate can be formed as a result of interaction of two quantum fields (as, for example, it occurs in lasers), not only as a result of cooling below the Bose-condensation temperature or the influence of external coherent field.

A property of the condensate to decay into various number of quanta results in essential broadening of spectrum of the SCLP photodissociation in comparison with that predicted by theories of Pekar [1] and Emin [6] with classical consideration of the polarization field. The band calculated above with quantum consideration of the polarization field is in good conformity with mid-infrared bands observed in optical conductivity spectra of different complex oxides. This confirms that decay of phonon condensate into phonons occurs at the SCLP photodissociation. In accordance with Pekar [1] prediction there are two bands in the optical conductivity spectrum of SCLP. One of them is caused by SCLP photodissociation, its maximum is at about $4.2E_p$, and the other is due to transitions into internal excited states. The latter is calculated in [5], its maximum is at about $E_p$. Relative positions of these maximums predicted theoretically are in good conformity with those observed experimentally [14, 15].

## Appendix 1. Calculation of the sum in Exp.(23).

To calculate the sum in Exp.(23) let us estimate the values of $|d_\mathbf{k}|^2$. Obviously, the number of summands in the sum $\sum_\mathbf{k} |d_\mathbf{k}|^2$ is determined by the size of Landau-Pekar polaron. Its volume can be hundreds times larger than the volume of the elementary cell of a crystal. In a crystal with $N$ crystal cells the number of summands in the mentioned sum is hundreds times smaller than $N$ and, hence, is of the order of $10^{20}$. Therefore the value of each summand $|d_\mathbf{k}|^2$ is very small. Indeed, estimation based on Exp. (5) at $m^* = m_e$ and $\varepsilon^* = 3$ shows that $|d_\mathbf{k}|$ is ordinarily of the order of $10^{-10}$. Only for the smallest values of $k \approx \pi a^{-1} N^{-1/3}$ the value $|d_\mathbf{k}|$ achieves $10^{-3}$.

Hence the probabilities $|\langle v | d_\mathbf{k} \rangle|^2$ determined by Exp.(24) are practically different from zero only at $v = 0$ and $v = 1$, and we can use the approximation

$$\left|\langle 0|d_{\mathbf{k}}\rangle\right|^{2} = \exp\{-|d_{\mathbf{k}}|^{2}\} \approx 1 - |d_{\mathbf{k}}|^{2}, \quad (A1)$$

$$\left|\langle 1|d_{\mathbf{k}}\rangle\right|^{2} = |d_{\mathbf{k}}|^{2} \exp\{-|d_{\mathbf{k}}|^{2}\} \approx |d_{\mathbf{k}}|^{2}. \quad (A2)$$

Obviously, the average value of quanta radiated by a polaron after its sudden photoionization turns out to be much larger than unity only due to the participation of a large number of harmonics in the process.

First let us consider a case of $\nu = 1$ with taking into account Exps.(A1, A2):

$$P_1(\mathbf{Q}-\mathbf{k}) \equiv \sum_{\{\nu_{\mathbf{q}}\}}^{1} \prod_{\mathbf{q}} \left|\langle \nu_{\mathbf{q}}|d_{\mathbf{q}}\rangle\right|^{2} = |d_{\mathbf{Q}-\mathbf{k}}|^{2} \prod_{\mathbf{q}\neq\mathbf{Q}-\mathbf{k}} \exp\{-|d_{\mathbf{q}}|^{2}\}. \quad (A3)$$

As only one factor in the product over $\mathbf{q}$ is absent and as all $|d_{\mathbf{q}}|^{2} \ll 1$ then, using (A1, A2), we can write with a good accuracy

$$P_1(\mathbf{Q}-\mathbf{k}) = |d_{\mathbf{Q}-\mathbf{k}}|^{2} \exp\{-\bar{\nu}\} = |d_{\mathbf{Q}-\mathbf{k}}|^{2} e^{-\frac{2E_p}{\hbar\omega}}. \quad (A4)$$

In the case of $\nu = 2$

$$P_2(\mathbf{Q}-\mathbf{k}) = \sum_{\{\nu_{\mathbf{q}}\}}^{2} \prod_{\mathbf{q}} \left|\langle \nu_{\mathbf{q}}|d_{\mathbf{q}}\rangle\right|^{2} = \sum_{\mathbf{q}_1} |d_{\mathbf{q}_1}|^{2} |d_{\mathbf{Q}-\mathbf{k}-\mathbf{q}_1}|^{2} \prod_{\mathbf{q}\neq\mathbf{Q}-\mathbf{k}-\mathbf{q}_1,\mathbf{q}_1} \exp\{-|d_{\mathbf{q}}|^{2}\}. \quad (A5)$$

It can be estimated again with a good accuracy as

$$P_2(\mathbf{Q}-\mathbf{k}) = \sum_{\mathbf{q}} |d_{\mathbf{q}}|^{2} |d_{\mathbf{Q}-\mathbf{k}-\mathbf{q}}|^{2} \exp\{-\bar{\nu}\}. \quad (A6)$$

If $(\mathbf{Q}-\mathbf{k})$ value is small in comparison with the width of the region where $\mathbf{q}$ changes the dependence of Exp.(A6) on $\mathbf{Q}-\mathbf{k}$ is weak.

As $\bar{\nu}$ and all the values of $\nu$ having essential probability are much smaller than the number of summands in the sums over $\mathbf{q}$, then the structure of the expression for $P_\nu(\mathbf{Q}-\mathbf{k})$ with any $\nu$ does not differ from (A6):

$$P_\nu(\mathbf{Q}-\mathbf{k}) = \sum_{\{\nu_{\mathbf{q}}\}}^{\nu} \prod_{\mathbf{q}} \left|\langle \nu_{\mathbf{q}}|d_{\mathbf{q}}\rangle\right|^{2} \cong \sum_{\mathbf{q}_1 \neq \ldots \neq \mathbf{q}_{\nu-1}} |d_{\mathbf{q}_1}|^{2} |d_{\mathbf{q}_2}|^{2} \ldots |d_{\mathbf{q}_{\nu-1}}|^{2} \left|d_{\mathbf{Q}-\mathbf{k}-\sum_{i=1}^{\nu-1}\mathbf{q}_i}\right|^{2} \exp(-\bar{\nu}). \quad (A7)$$

In the cases of $\nu > 2$ we can neglect the dependence of $P_\nu(\mathbf{Q}-\mathbf{k})$ on $\mathbf{Q}-\mathbf{k}$ because of a large number of factors in the product that do not depend on $\mathbf{Q}-\mathbf{k}$.

The number of summands in the sum in (A7) is, obviously, the number of combinations $C_N^{\nu-1}$, where N is the number of possible values of the wave vector. As $N \gg \nu$, $C_N^{\nu-1} \approx \frac{N^{\nu-1}}{(\nu-1)!}$. Let us introduce a value $\overline{|d|^{2}}$ as it follows:

$$N\overline{|d|^{2}} = \sum_{\vec{q}} |d_{\vec{q}}|^{2} = \bar{\nu}. \quad (A8)$$

Then the value (A7) can be estimated as

$$P_\nu \approx \frac{\left(N\overline{|d|^{2}}\right)^{\nu}}{(\nu-1)!} \cdot \frac{e^{-\bar{\nu}}}{N} = \frac{(\bar{\nu})^{\nu}}{(\nu-1)!} \cdot \frac{e^{-\bar{\nu}}}{N} = \frac{\bar{\nu}^{\nu-1}}{(\nu-1)!} \frac{\bar{\nu} e^{-\bar{\nu}}}{N}. \quad (A9)$$

Obviously, a sum of $P_\nu$ over $\nu$ should be unity as a sum probability of all possible decays of the phonon condensate upon the polaron photodissociation. However, summation of (A9) over $\nu$ does not yield 1 since we do not take into account in (A9) the uncertainty of the momentum of the phonon field in SCLP. In order to obtain physical characteristics of spectrum in absolute units we will normalize $P_\nu$ as follows:

$$P_\nu = \frac{\bar{\nu}^{\nu-1}}{(\nu-1)!} e^{-\bar{\nu}} \qquad (A10)$$

The normalization compensates the lack of information caused by using the adiabatic approximation.